\documentstyle[12pt]{article}

\newcounter{ctr}

\begin{document}
\renewcommand{\thefootnote}{\fnsymbol{footnote}}
\begin{titlepage}

\vspace*{-15mm}
\begin{flushright}
\normalsize{BAIKA-93-01}
\end{flushright}
\vspace{8mm}
\begin{center} {\large\bf Characteristic Representation of
Elementary}\\{\large\bf Cellular Automata\footnote[1] {Work supported in
part
by BAIKA  Women's College, Japan.}}\\ \vspace{12mm} {\normalsize Yoshihiko
KAYAMA}\\ {\small\em Department of General Education, BAIKA Women's
College}\\  {\small\em Shukuno-sho 2-19-5, Ibaraki-shi, Osaka 567, Japan}\\

and\\ {\normalsize Hajime ANADA}\\ {\small\em Department of Earth and
Planetary Sciences}\\ {\small\em Kobe University, Nada, Kobe 657, Japan.}\\
and\\ {\normalsize Yasumasa IMAMURA}\\{\small\em  The Graduate School of
Science and Technology}\\ {\small\em Kobe University, Nada, Kobe 657,
Japan.}\\ \vspace{4mm} \normalsize October 1993\\ \end{center}

 \begin{abstract} We propose a characteristic representation of
one-dimensional and 2-state, 3-neighbor cellular automaton rules, which
describes an effective form of each rule after many time steps. Simulated
results of the representation show that complex structures of Class
\setcounter{ctr}{4}\Roman{ctr} rules come from their aspects of Class
\setcounter{ctr}{2}\Roman{ctr} and \setcounter{ctr}{3}\Roman{ctr}. Class
\setcounter{ctr}{4}\Roman{ctr}-like patterns can be generated by
characteristic functions of a linear combenation of Class
\setcounter{ctr}{2}\Roman{ctr} and \setcounter{ctr}{3}\Roman{ctr} rule
functions. \end{abstract}  \newpage

\end{titlepage}

\newenvironment{indention}[1]{\par
\addtolength{\leftskip}{#1} \begingroup}{\endgroup\par}

\setcounter{equation}{0}

\section{Introduction}

Cellular automata (CAs) are discrete dynamical systems with a lattice of
sites. Each site has a finite set of possible values which evolve
synchronously in discrete time steps according to identical rules.
Wolfram\cite{wolf83}\cite{wolf84} undertook valuable works to search
systematically for interesting phenomena and possible behavior of
one-dimensional automata.  He found essential four types of behavior:
homogeneous (Class \setcounter{ctr}{1}\Roman{ctr}), periodic (Class
\setcounter{ctr}{2}\Roman{ctr}), chaotic (Class
\setcounter{ctr}{3}\Roman{ctr}) and complex (Class
\setcounter{ctr}{4}\Roman{ctr}). This classification scheme has been
discussed using several characteristic parameters[3-11].  In the previous
work\cite{kaya93} of one of the authors and collaborators, characteristic
parameters are calculated in all rules of the simplest one-dimensional CAs
and applied to give quantitative grounds  for the Wolfram's
phenomenological
classification.  The simplest CAs consist of a line of sites with each site
carrying a value 0 or 1.  The values of a particular site is determined by
the previous values of its nearest neighbors. They are called
one-dimensional
and $2$-state, $3$-neighbor CAs or $elementary$ CAs. Rules $54$  and $110$,
following Wolfram's  numbering\cite{wolf84},  are thought to be in Class
\setcounter{ctr}{4}\Roman{ctr}\cite{li90}. Typical patterns generated by
them
show long transients and soliton-like phenomena(Fig. 1) .  According to the
previous paper\cite{kaya93}, $D_{iff}$, the spreading rate of difference
patterns, is the most powerful parameter to classify the rules.  The
calculated results are nearly sufficient for categorizing the rules to
Class
\setcounter{ctr}{1}\Roman{ctr}, \setcounter{ctr}{2}\Roman{ctr} and
\setcounter{ctr}{3}\Roman{ctr}. The Class \setcounter{ctr}{4}\Roman{ctr}
rules, however, could not be distinguished from Class
\setcounter{ctr}{3}\Roman{ctr} rules.  Their results of $D_{iff}$ are
tentatively sorted in Table 1.  It is remarkable that rules $54$ and $110$
exist between Class \setcounter{ctr}{2}\Roman{ctr} and
\setcounter{ctr}{3}\Roman{ctr} groups.  This result supports that the
formalism is available for characterizing Class
\setcounter{ctr}{4}\Roman{ctr} rules and we can conjecture that their
features come from interference between  their aspects of Class
\setcounter{ctr}{2}\Roman{ctr} and \setcounter{ctr}{3}\Roman{ctr}.
Simultaneously, it represents the confines of discussions on a single
characteristic parameter. Because rule $41$, for example, has the close
value
of $D_{iff}$ with those of rules $54$ and $110$, we cannot distinguish
them.
So we present a characteristic representation of each rule, which is given
by
a linear combination of independent rule functions.  We can use this
formalism to classify CA rules because it describes an effective form of
each
rule function after many time steps.


\section{Characteristic Representation} \label{sec:form} We take $x_k$ to
denote a value of the $k$th site in a one-dimensional Cellular Automaton.
Each site value is specified as an integer $0$ or $1$ ($2$-state). If there
are $N$ total sites, a configuration can be denoted by  \begin{equation}
X=(x_1,x_2,\ldots,x_N)\ . \end{equation} The Hamming distance of $X$ from
${\bf 0}=(0,\ldots,0)$ is identical to its $weight$, \begin{eqnarray} w(X)
=
\sum_{k=1}^{N}{x_k}\ . \label{eqn:energy} \end{eqnarray} We define a
configuration space $\left\{ X \right\}_N$ which is a set of all distinct
configurations represented by the above form.  Time evolution of a site
value
is determined by iteration of the mapping \begin{eqnarray} x_{k}^{(t+1)} &
=
&f(x_{k-r}^{(t)},\ldots,x_{k}^{(t)},\ldots,x_{k+r}^{(t)}) \nonumber\\ &
\equiv & f_{r}(x_{k}^{(t)}), \label{eqn:rule} \end{eqnarray} where $f$ is
called a $rule$ $function$ which specifies the CA.  The subscript $r$
determines a neighborhood of the rule  function ($(2r+1)$-neighbor). In the
following, we abbreviate the subscript because our discussions are
restricted
to $3$-neighbor CAs. A rule function is determined by the succeeding $8$
values \begin{equation} f(0,0,0),f(0,0,1),\cdots ,f(1,1,1), \end{equation}
or
$f(0),f(1),\cdots,f(7)$ for short. There are $2^8=256$ rules of the
elementary CAs and rule numbers are defined by \begin{equation}
\sum\limits_{i=0}^7 {f(i)\cdot 2^i}. \end{equation} Following the notation
of
the previous paper\cite{kaya93}, if $f^{(1)}$ denotes a rule function of
the
first time step, the function of the $t$th time step can be obtained by
\begin{equation} f^{(t)}(x_k)  =  f^{(1)}(f^{(1)}(\cdots
f^{(1)}(x_k)\cdots))\nonumber\\, \end{equation} which is a rule for
$(2t+1)$-neighbor of $x_k$: \begin{equation}
f^{(t)}(x_k)=f^{(t)}(x_{k-t},\ldots,x_{k},\ldots,x_{k+t})\ . \end{equation}
By the $t$th rule function we can define a mapping on the configuration
space
$\left\{ X \right\}_N$, $\mbox{\boldmath$f$}^{(t)} :\left\{ X \right\}_N
\to
\left\{ X \right\}_N$ such as \begin{eqnarray} \mbox{\boldmath$f$}^{(t)}(X)
&
= & \mbox{\boldmath$f$}^{(t)}((x_1,x_2,\ldots,x_N))\nonumber\\ & = &
(f^{(t)}(x_1),f^{(t)}(x_2),\ldots,f^{(t)}(x_N)), \end{eqnarray} with a
proper
boundary treatment. In this article, we adopt the periodic boundary
condition. The image of  $\mbox{\boldmath$f$}^{(t)}$,
$\mbox{\boldmath$f$}^{(t)}(\left\{ X \right\}_N)$, is a subset of $\left\{
X
\right\}_N$.   Since the following is a monotone decreasing sequence of
$\mbox{\boldmath$f$}^{(t)}(\left\{ X \right\}_N)$'s, \begin{equation}
\left\{X\right\}_N\supset \mbox{\boldmath$f$}^{(1)}(\left\{X \right\}_N)
\supset \mbox{\boldmath$f$}^{(2)}(\left\{ X \right\}_N)\supset \cdots
\supset
\mbox{\boldmath$f$}^{(t)}(\left\{ X \right\}_N)\supset \cdots\ ,
\label{eqn:imgseq} \end{equation} there exists a limit set \small
${\displaystyle\lim_{\scriptscriptstyle{t\to\infty}}}$
\normalsize$\mbox{\boldmath$f$}^{(t)}(\left\{ X \right\}_N )$, which will
be
thought to contain characteristics of each CA because its asymptotic
behavior
is fixed to an element or a subset of this set.  Classification of CA rules
will be carried out through arrangement of rule functions by similarity of
their limit sets.  If limit sets of two different rules are equivalent or
similar for sufficiently large total sites,  we can recognize these rule
functions to be effectively the same, which suggests the existence of a
representation of rule functions taking the similarity of limit sets into
account.

We take $f_{R}^{(1)}$ to denote the rule function of rule $R$ of the
elementary CAs.  After $t$ time evolutions, the domain of the mapping
$\mbox{\boldmath$f$}_{R}^{(1)}$  to the $(t+1)$th time step is
$\mbox{\boldmath$f$}_{R}^{(t)}(\left\{ X \right\}_N)$.  Because
$\mbox{\boldmath$f$}_{R}^{(t)}:\left\{ X \right\}_N \to \left\{ X
\right\}_N$
is a many-to-one mapping in general,  we define multiplicity of the mapping
for an element $X^{(t)}$ of $\mbox{\boldmath$f$}_{R}^{(t)}(\left\{ X
\right\}_N)$ as $m_R\left( {X^{(t)}} \right)$. Moreover $n_R\left(
{i,X^{(t)}} \right)$ denotes how many times each configuration of three
sites, $(0,0,0),(0,0,1),\cdots,(1,1,1)$ ($i=0,1,\ldots,7$, respectively),
appears in a configuration $X^{(t)}$. Then the following equations are
trivial: \begin{equation} \sum\limits_{X^{(t)}\in f_{R}^{(t)}\left( {{X}_N}
\right)} {m_R\left( {X^{(t)}} \right)}=2^N,\ \ \ \  \sum\limits_{i=0}^7
{n_R\left( i,{X^{(t)}} \right)}=N. \end{equation} Using these parameters,
the
appearance probability of each configuration at the $t$th time step is
obtained by \begin{equation} P_{R}^{(t)}\left( i \right)={1 \over {2^N\cdot
N}}\sum\limits_{X^{(t)}\in f_{R}^{(t)}\left( {\left\{ X \right\}_N}
\right)}
{m_R\left( {X^{(t)}} \right)n_R\left( i,{X^{(t)}} \right)}. \end{equation}
Then the average weight of a configuration $X^{(t+1)}$ is given by
\begin{equation} \left\langle {w_{R}\left( {X^{(t+1)}} \right)}
\right\rangle_N =N\sum\limits_{i=0}^7 {P_{R}^{(t)}\left( i
\right)}f_{R}^{(1)}\left( i \right). \end{equation}

Modifying the rule function $f_{R}^{(1)}$, we define the $t$th
$characteristic$ $function$ $\tilde f_{R}^{(t)}$.  The following conditions
are imposed on it: \begin{indention}{0.5cm} \noindent (i) The mapping
$\tilde
{\mbox{\boldmath$f$}}_{R}^{(t)}$ constructed from $\tilde f_{R}^{(t)}$ is
defined on $\left\{ X \right\}_N$. \\(ii) The value of a characteristic
function is limited between 0 and 1: \begin{equation} 0\le \tilde
f_{R}^{(t)}\left( i \right)\le 1\ \ \ \ \mbox{for}\ \ i=0,1,\ldots ,7.
\end{equation} (iii) The average weight of a configuration contained in the
image of the mapping $\tilde {\mbox{\boldmath$f$}}_{R}^{(t)}$ is equal to
the
above average weight of $X^{(t+1)}$. \end{indention} \noindent The first
condition means that the appearance probability $P_{R}\left( i \right)$ is
equal to ${\textstyle{1 \over 8}}$.  According to the second, a
characteristic function cannot remain in CAs anymore.  The image of $\tilde
{\mbox{\boldmath$f$}}_{R}^{(t)}$ is a subset of the unit $N$-cube. As we
discuss in later, a probability interpretation must be given. Although the
images $\mbox{\boldmath$f$}_{R}^{(t+1)}(\left\{ X \right\}_N)$ and $\tilde
{\mbox{\boldmath$f$}}_{R}^{(t)}(\left\{ X \right\}_N)$ are generally not
identical, we must define $\tilde f_{R}^{(t)}$ as the third condition is
satisfied. From these conditions, we have \begin{equation} \left\langle
{w_{R}\left( {X^{(t+1)}} \right)} \right\rangle_N ={N \over
8}\sum\limits_{i=0}^7 {\tilde f_{R}^{(t)}}\left( i \right). \end{equation}
Now we propose the following definition of a characteristic function:
\begin{equation} \tilde f_{R}^{(t)}(i)=\left\{ {\matrix{{\ \ \ \ \ \ \ \ \
\
f_{R}(i)\ \ \ \ \ \ \ \ \ \ \ \ \ \ \ \ \ \ \ \mbox{for}\ \ \ P_{R}(i)\ge
{1
\over 8}},\cr \matrix{\hfill\cr 8P_{R}(i)f_{R}(i)+{{\sum\limits_{j\
\mbox{\tiny for}\ P_{R}(j)\ge {1 \over 8}} {\left( {P_{R}(j)-{1 \over 8}}
\right)f_{R}(j)}} \over {\sum\limits_{j\ \mbox{\tiny for}\ P_{R}(j)\ge {1
\over 8}} {\left( {P_{R}(j)-{1 \over 8}} \right)}}}\left( {1-8P_{R}(i)}
\right)\hfill\cr}\cr {\ \ \ \ \ \ \ \ \ \ \ \ \ \ \ \ \ \ \ \ \ \ \ \ \ \ \
\
\ \ \ \ \ \ \ \ \ \mbox{for}\ \ \ P_{R}(i)<{1 \over 8}.}\cr }} \right.
\end{equation} When $t=0$, $\tilde f_{R}^{(0)}$ is identical with the rule
function $f_{R}^{(1)}$ because there is no effect of the time evolution.
Let
us define a function space $\left\{ {\tilde f|\tilde f:\left\{ X
\right\}_3\to [0,1]} \right\}$. Then the characteristic function and all
rules of the elementary CAs, from rule $0$ to rule $255$, are contained in
it.  If a product of elements are defined by \begin{equation} \left(
{\tilde
f\bullet \tilde g} \right)\equiv \prod\limits_{i=0}^7 {\left( {\tilde
f\left(
i \right)\cdot \tilde g\left( i \right)+( {1-\tilde f\left( i \right)}
)\cdot
\left( {1-\tilde g\left( i \right)} \right)} \right)}, \end{equation} the
characteristic function can be expanded by the elementary CA rule functions
as follows: \begin{equation} \tilde f_{R}^{(t)}=\sum\limits_{r=0}^{255}
{\left( {f_{r}\bullet \tilde f_{R}^{(t)}} \right)}f_{r}. \end{equation} The
coefficient $\left( {f_{r}\bullet \tilde f_{R}^{(t)}} \right)$ means a rate
of change from the original rule $R$ to  rule $r$ after $t$ time
evolutions.
We call this expanded form as a $characteristic$ $representation$ of the
rule
function.  In order to show adequacy of this representation, we calculated
the coefficients for all independent even number rules of the elementary
CAs
by computer simulation. Some typical examples are listed in Table 2. All
Class \setcounter{ctr}{1}\Roman{ctr} rules change to rule $0$. Most of
Class
\setcounter{ctr}{2}\Roman{ctr} rules change to itself or typical Class
\setcounter{ctr}{2}\Roman{ctr} rules, say $2$, $4$, $34$.  Class
\setcounter{ctr}{3}\Roman{ctr} rules almost stay in the same Class.  It is
remarkable, however, that the Class \setcounter{ctr}{4}\Roman{ctr} rules
$54$
and $110$ have larger changing rate to Class \setcounter{ctr}{2}\Roman{ctr}
rules than Class \setcounter{ctr}{3}\Roman{ctr} rules (Table 3).  These
results indicate that their complex structures come from the interaction
between aspects of Class \setcounter{ctr}{2}\Roman{ctr} and
\setcounter{ctr}{3}\Roman{ctr} rules.  Repetitions of limited patterns as
the
aspect of Class \setcounter{ctr}{2}\Roman{ctr} are affected by chaotic
variations as the aspect of Class \setcounter{ctr}{3}\Roman{ctr}, and then
complex patterns, soliton-like behaviors for example, are generated. In
fact,
as we will show in the next section, such Class
\setcounter{ctr}{4}\Roman{ctr}-like patterns can be generated by
characteristic functions of a linear combination of Class
\setcounter{ctr}{2}\Roman{ctr} and \setcounter{ctr}{3}\Roman{ctr} rule
functions. Now we conclude that the Class \setcounter{ctr}{4}\Roman{ctr}
rules exist between Class \setcounter{ctr}{2}\Roman{ctr} and
\setcounter{ctr}{3}\Roman{ctr} rules, that is, the so-called $edge$ $of$
$chaos$.


\section{Discussion} Because the $t$th rule function $f^{(t)}$ of the
elementary CAs is determined by $2^{2t+1}$ values,  it is difficult to
obtain
its exact algebraic form after many time steps in general.  Instead, we
considered the characteristic representation which took into account the
effects of the time evolution  not directly but through changes of the
appearance probabilities of the $3$ sites configurations.   It has an
expansion form by all rule functions so that each coefficient means a
changing rate from the original rule to the corresponding one.  This
representation makes clear features of Class \setcounter{ctr}{4}\Roman{ctr}
rules of the elementary CAs as well as can be used to classify the CA
rules.
As noted in Section 2, the calculated results showed that the changing
rates
from Class \setcounter{ctr}{4}\Roman{ctr} rules to Class
\setcounter{ctr}{2}\Roman{ctr} rules are larger than the rates from Class
\setcounter{ctr}{3}\Roman{ctr} rules to Class
\setcounter{ctr}{2}\Roman{ctr}
rules.  Using the characteristic representations, we can generate
interference patterns such as Class \setcounter{ctr}{4}\Roman{ctr} rules.
If
we regard our representation as an extension of CAs to a stochastic model,
each value of the characteristic function $\tilde f(x,y,z)$ can be
interpreted as the probability that the site $y$ will have a value $1$ at
the
next time step. For example, we take such functions that \begin{equation}
\tilde f={1 \over 2}f_{4}+{1 \over 2}f_{90},\ \ \ \ \ \tilde g={1 \over
2}f_{232}+{1 \over 2}f_{90}, \end{equation} where rules $4$, $232$ and $90$
are typical rules of Class \setcounter{ctr}{2}\Roman{ctr} and
\setcounter{ctr}{3}\Roman{ctr}, respectively.  These functions generate
complex patterns as Fig. 2 which are analogous to the patterns of Class
\setcounter{ctr}{4}\Roman{ctr} rules. Such interference patterns support
the
conjecture presented in Section 1, that is to say, features of Class
\setcounter{ctr}{4}\Roman{ctr} rules comes from interference between the
aspects of Class \setcounter{ctr}{2}\Roman{ctr} and
\setcounter{ctr}{3}\Roman{ctr} rules.

Away from the theme of the classification of CAs, this dynamical system is
very interesting. A pattern construction process can be shown by a
continuous
change of proper elements of $\tilde f(x,y,z)$. For example, rule $46$
changes to rule $110$ by varying the value of $\tilde f(1,1,0)$ from $0$ to
$1$(Fig. 3). Moreover, this formalism will be able to be applied to more
complex systems such as $5$-neighbor CAs (now in preparation).


\newpage

\begin{thebibliography}{99} \bibitem{wolf83}S. Wolfram, Statistical
mechanics
of cellular automata, $Rev.$ $Mod.$ $Phys.$ 55 (1983) 601.
\bibitem{wolf84}S.
Wolfram, Universality and complexity in cellular automata, $Physica$ D 10
(1984) 1. \bibitem{gras84}P. Grassberger, Chaos and diffusion in
deterministic cellular automata, $Physica$ D 10 (1984) 52.
\bibitem{wolfbook}S. Wolfram (ed.), $Theory$ $and$ $Applications$ $of$
$Cellular$ $Automata$, World Scientific, Singapore (1986).
\bibitem{gras83}P.
Grassberger, New mechanism for deterministic diffusion, $Phys.$ $Rev.$ A 28
(1983) 3666. \bibitem{li90}W. Li and N. Packard, The Structure of the
Elementary Cellular Automata Rule Space, $Complex$ $Systems$ 4 (1990) 281.
\bibitem{lipacka90}W. Li, N. Packard and C. Langton, Transition phenomena
in
cellular automata rule space, $Physica$ D 45 (1990) 77.
\bibitem{lanton90}C.
Langton, Computation at the edge of chaos: phase transitions and emergent
computation, $Physica$ D 42 (1990) 12. \bibitem{gras86}P. Grassberger,
Long-range effects in an elementary cellular automaton, $J.$ $Stat.$
$Phys.$
45 (1986) 27.   \bibitem{packa85}N. H. Packard, Complexity of growing
patterns in cellular automata, in $Dynamical$ $systems$ $and$ $cellular$
$automata$, edited by J. Demongeot, E. Goles and M.Tchuente, Academic
Press,
New York (1985).  \bibitem{kaya93}Y. Kayama, M. Tabuse, H. Nishimura and T.
Horiguchi, Characteristic Parameters and Classification of One-dimensional
Cellular Automata, $Chaos,$ $Solitons$ $\&$ $Fractals$ 3 (1993), to be
published. \end{thebibliography}
 \end{document}